# The Complementary Betz's Theory

Adriano Pellegri

Correspondence to: A. Pellegri (adriano_pellegri@yahoo.it)

**Abstract.** A classical derivation of Betz's law is first presented along with some insights. The extended Betz's theory is deduced for a rotor with axis *orthogonal* to the direction of an ideal fluid in uniform motion. The conceptual design used to demonstrate the generality of the aerodynamic aspects of energy conversion – starting from a suggestive explanation of Betz's law – is defined by imposing compliance with the minimum requirements that a turbine must have to approximate the ideal efficiency of 16/27. The discovery of the role of the *Betz's angles* has permitted demonstrating the theory in a more general context, extending it from a flat, two-dimensional, representation to a deeper, three-dimensional, understanding of the problem. The ensuing result, in particular the possibility of obtaining constant yield and maximum efficiency through a finite topology of the *actuator cylinder*, implies in principle feasibility with a real turbine with axis perpendicular to the wind direction.

## 1 Introduction

The Betz-Joukowsky limit (16/27 ≈ 59.3%), and the condition on the ideal ratio (3:1) between inflow and outflow speeds, as we know, can be deduced using the so-called *actuator disk model*, where the turbine is constituted by a thin disk, free to rotate about its axis, through which the fluid flows in axial direction.

The deduction of Betz's law rests on energy balance considerations applicable to all wind turbines, with either horizontal axis (HAWT) or, as we shall prove analytically, vertical axis (VAWT), employing a substitute of the actuator disk. It has been generally stated that the actuator disk theory, although useful for establishing the limit of efficiency, does not help in designing high performance real turbines. In this paper, we shall see how this statement needs reviewing.

The explicit reference in Betz's work[1] to a rotating disk with horizontal axis, combined with the fact that, notoriously, the HAWTs can achieve performances of 40–45% against 20–30% of the VAWTs, has strengthened over the years the belief that high efficiencies are only possible with turbines with axis parallel to the wind direction and has rooted in the wind energy sector the idea that the three-blade mills represent today the technological limit of our ability to produce energy from the wind, and we can only aim to reduce the costs of construction and maintenance of a wind farm by realizing ever larger mills.

The theoretical maximum efficiency value is derived from the principles of conservation of mass and energy of a stream flowing through an ideal energy converter, able to extract energy from the fluid by reducing its velocity module. An important but often neglected aspect of Betz's theory is that it also shows that this limit is not absolute. The Betz limit depends in some way on geometry. Indeed, even the idealized turbine cannot extract the maximum energy from the fluid if the speed ratio between outgoing and incoming streams is not exactly one third, the so-called Betz condition:

$$\frac{v_{out}}{v_{in}} = \frac{1}{3}$$

In this article, a derivation of Betz's law is first presented along with some insights (§2). We then introduce (§3) a suggestive *geometrical* explanation of the classical theory, which provides a new approach to the axial momentum theory and leads to the definition of the *actuator cylinder* concept (§4) and hence to the complementary, extended, Betz's theory (§5). A closed analytical investigation of some aerodynamic aspects of the theory, through a simplified, finite, actuator cylinder model under ideal conditions, is introduced and developed (§6 and §7). Discussion and preliminary conclusions are finally presented (§8).

## 2 Betz's Law or Axial Momentum Theory

The relevant assumptions on which the Rankine-Froude actuator disk theory[2] is based are the following:

- ➢ *Steady, homogeneous flow*. Fluid velocity and pressure are not functions of time.
- ➢ *Incompressible flow*. Fluid density is constant.
- ➢ *Inviscid flow*. There is no resistance on rotor blades or heat transfer.
- ➢ *Uniform flow*. Fluid action is the same at any point of the rotor plane.
- ➢ *Irrotational wake*. No induced vorticity in the downstream flow.
- ➢ *Infinite number of blades* of infinitesimal chord length. The rotor is assumed to be an ideal energy converter.

Assuming an irrotational axial flow, the only effect of the turbine rotor on the flow which is expected is the variation of static pressure and axial momentum. A far away observer will see an increment of the static pressure upstream of the rotor, inside the stream tube, from the free stream pressure $p_0$ to a value $p^+ > p_0$, and a consequent reduction of the fluid speed, from its free stream value $U_0$ to a lower value $U$ at the rotor plane. The turbine will act as a porous obstacle producing a sudden local pressure drop, from $p^+$ to $p^- < p_0$, as the fluid passes through it. Downstream of the rotor the pressure increases up again to the undisturbed free stream value, and consequently the fluid speed will continue to decrease to a final value $U_1 < U < U_0$.

The velocity will not result in discontinuous variations, therefore the cross-sectional area of the stream tube increases from upstream of the turbine, $A_0$, to downstream of the turbine, $A_1$, and $A_0 < A < A_1$, where $A$ is the area swept by the rotor.

Indicating with $\rho$ the fluid density, from the conservation of mass or the continuity equation we know that the mass flow rate must be constant along the stream tube, thus:

$$\rho U_0 A_0 = \rho U A = \rho U_1 A_1 \tag{1}$$

To a faraway observer the thrust $T$ on disk will be equal to the axial momentum flow difference between the inlet section and the outlet section of the stream tube:

$$T = \rho U_0 A_0 U_0 - \rho U_1 A_1 U_1 = \rho U A (U_0 - U_1) \tag{2}$$

The force exerted by the fluid on the turbine rotor can also be expressed as the pressure gradient across the rotor disk:

$$T = \rho U A (U_0 - U_1) = A(p^+ - p^-) \tag{3}$$

It is not possible to integrate the Euler's equation of mechanical energy across the whole stream tube due to the discontinuity in the pressure function. We can separately apply Bernoulli's equation instead, *i*) between the inlet section of the stream tube and the section immediately upstream of the rotor disk, and *ii*) between the section immediately downstream and the outlet section of the stream tube:

$$\begin{cases} p_0 + \frac{1}{2}\rho U_0^2 = p^+ + \frac{1}{2}\rho U^2 \\ p^- + \frac{1}{2}\rho U^2 = p_0 + \frac{1}{2}\rho U_1^2 \end{cases} \tag{4}$$

which gives:

$$p^+ - p^- = \frac{1}{2}\rho(U_0^2 - U_1^2) \tag{5}$$

From equations (3) and (5) we have:

$$U(U_0 - U_1) = \frac{1}{2}(U_0 - U_1)(U_0 + U_1) \tag{6}$$

which implies that:

$$U = \frac{1}{2}(U_0 + U_1) \quad \forall \ U_0 \neq U_1 \tag{7}$$

The fluid velocity at the rotor may be regarded as the average of the upstream and downstream fluid velocities.

The actuator disk extracts energy from the fluid stream by reducing its velocity from $U_0$ to $U_1$, but not totally resting it down to $U = 0$, where the equations are no more valid: there must be some residual flow through the turbine, however small, and this poses an upper bound on the fraction of available energy which can be extracted, even under ideal conditions.

The only known velocity in the stream tube is the inlet free stream velocity $U_0$. Continuing with the derivation, following Froude we introduce two dimensionless parameters called the *axial induction coefficients*, $a$ and $b$, which set a relationship between the velocities in the various sections and the initial velocity:

$$U \equiv (1-a)U_0 \tag{8}$$

$$U_1 \equiv (1-b)U_0 \tag{9}$$

Using equation (7) for the average fluid velocity at the rotor, we can relate the induction in the rotor plane, $a$, with the induction in the wake, $b$:

$$(1-a)U_0 = \frac{1}{2}[U_0 + (1-b)U_0] \tag{10}$$

$$a = \frac{1}{2}b \tag{11}$$

We observe that the downstream velocity $U_1$ can at minimum vanish, therefore $0 < b \leq 1$ and $0 < a \leq 0.5$.

Using again equation (7), the thrust exerted on the disk expressed by (3) can be written as:

$$T = \frac{1}{2}\rho A(U_0^2 - U_1^2) \tag{12}$$

Power is the flow of energy, in our case the kinetic energy of the fluid. From the thrust $T$ and the velocity of the fluid at the rotor disk $U$ it is possible to calculate the power $P$ extracted by the turbine:

$$P = T \cdot U = \tfrac{1}{2}\rho A(U_0^2 - U_1^2)U \tag{13}$$

The total available power on a section equal to that swept by the actuator disk will be the product between the mass flow rate across the disk area and the kinetic energy associated to it:

$$P_{tot} = \tfrac{1}{2}\rho A U_0^3 \tag{14}$$

We can define the performance coefficient $C_p$ as the ratio of the extractable power (13) to the total kinetic power available in the undisturbed stream (14):

$$C_p \equiv \frac{P}{P_{tot}} = \frac{U}{U_0}\left(1 - \frac{U_1^2}{U_0^2}\right) \tag{15}$$

The performance coefficient is a dimensionless measure of the efficiency of the turbine in extracting the energy content of the stream. It is possible to express it as a function of the induction coefficients, $a$ or $b$:

$$C_p(a) = (1 - a)[1 - (1 - b)^2] = 4a(1 - a)^2 \tag{17a}$$

$$C_p(b) = (1 - a)[1 - (1 - b)^2] = \tfrac{1}{2}b(2 - b)^2 \tag{17b}$$

where we used the relation (11) between the two induction coefficients.

The condition for maximum efficiency is obtained by differentiating either (17a) or (17b), setting the derivative equal to zero and solving for $a$ or $b$, respectively:

$$\begin{cases} \frac{dC_p(a)}{da} = 4(1 - a)(1 - 3a) = 0 \\ \frac{dC_p(b)}{db} = (2 - b)(1 - \tfrac{3}{2}b) = 0 \end{cases} \tag{18}$$

which both yield the practical physical solution known as the Lanchester-Betz-Joukowsky limit:

$$\boxed{a = \tfrac{1}{3} \quad U = \tfrac{2}{3}U_0} \tag{19a}$$

$$\boxed{b = \tfrac{2}{3} \quad U_1 = \tfrac{1}{3}U_0} \tag{19b}$$

$$\boxed{C_{p,max} = \tfrac{16}{27} = 0.592593} \tag{19c}$$

# 3 A geometrical proof of Betz's law

In the classical derivation of the theory, the ideal rotor is taken at rest and uniformity is assumed over the whole area swept by the actuator disk. The aim of that simple mathematical model is to obtain a first order estimate of the wake-induced flow and total power loss. One of the main assumptions is that the actuator disk has a surface of zero thickness and no drag, through which the fluid experiences an instantaneous negative pressure gradient and a continuous deceleration. The discontinuity is removed by considering the wind velocity at the disk as the arithmetic average of the free stream speed and the downstream speed, arguably applying Bernoulli across the flow discontinuity.

Physically, the disk represents a rotor with a countless number of infinitely thin blades. The actuator disk model is thus an approximation of a real wind turbine not only because the latter has a small number of blades but mainly because the actuator disk allows a uniform thrust loading. Indeed, uniformity requires that the disk must slow the fluid equally at each radius (annular independency), which is equivalent to assume an infinite number of locally parallel rotor blades, as we shall see.

Blade Element Momentum model is today the industrial standard employed for many wind turbines design and analysis purposes. A basic BEM implementation relies on several assumptions originally introduced by Glauert[3], in order to simplify the problem to a suitable computational level[4, 5], such as annular independency and constant induction over the rotor plane equal to half of the wake induction, that is, Froude's result for optimal efficiency (11). Moreover, in the axial momentum balance, the pressure gradient between the rotor plane and the wake plane is ignored, as well as a tip loss factor is introduced to account for the difference between a finite-bladed rotor and the actuator disk on which the BEM method is based[6].

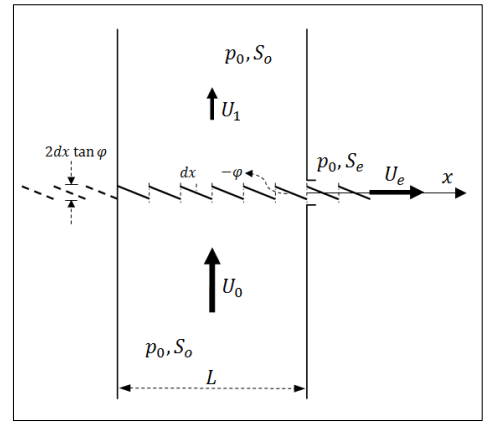

**Figure 1. Sliding energy extractor.**

Therefore, it is interesting the fact that we can achieve the same result of the axial momentum theory of Betz with a slightly different approach which allows considering less stringent assumptions by introducing a **geometric factor**.

To ease the derivation for a rotating disk, which we are most interested in, we start with a simpler model where the ideal energy extractor is constituted by a sliding planar surface, orthogonal to the flow, constrained to move without friction parallel to itself along the $x$-axis. As shown in Fig. 1, the energy extractor surface has a fine structure composed of an infinite number of stripes of height $H$, length $2dx/\cos\varphi$, each one inclined with respect to the plane of an angle $-\varphi$. The considered rectangular stream tube has a constant cross-sectional area $S_0 = L \cdot H$. Given the same assumptions of the actuator disk, that is, for a steady, incompressible, uniform, axial flow, a sudden change in the axial velocity of the flow is possible without the necessity to introduce a pressure gradient, if we further assume that the energy extractor absorbs the kinetic energy lost by the fluid through an ideal drag effect, where the zero-inertia extractor moves with the same velocity of the mass-loss flux (inelastic collision). We can easily verify the following relations between the free stream speed $U_0$, the downstream flow speed $U_1$ and the extractor speed $U_e$:

$$U_e = U_0|\cos\varphi| \equiv (1 - a')U_0 \tag{20a}$$

$$U_1 = U_0|\sin\varphi| \equiv (1 - b)U_0 \tag{20b}$$

Being the pressure $p_0$ constant at every point of the considered stream tube we have that Bernoulli is immediately satisfied:

$$\tfrac{1}{2}\rho(U_e^2 + U_1^2) = \tfrac{1}{2}\rho U_0^2(\cos^2\varphi + \sin^2\varphi) = \tfrac{1}{2}\rho U_0^2 \tag{21}$$

From the continuity equation, because the wake does not expand, we get:

$$\rho U_0 S_0 = \rho(U_e S_e + U_1 S_0) = \rho U_0(S_e|\cos\varphi| + S_0|\sin\varphi|) \tag{22}$$

hence, for $\cos\varphi \neq 0$,

$$S_e = S_0 \frac{1 - |\sin\varphi|}{|\cos\varphi|} \tag{23}$$

By hypothesis, the power absorbed by the energy extractor is the kinetic energy loss per unit of time:

$$P = \tfrac{1}{2}\dot{m}U_e^2 = \tfrac{1}{2}\rho S_e \frac{dL}{dt}U_e^2 = \tfrac{1}{2}\rho S_e U_e^3 \tag{24}$$

Given the total available power, $P_{tot} = \tfrac{1}{2}\rho S_0 U_0^3$, the performance coefficient $C_p$ becomes:

$$C_p = \frac{P}{P_{tot}} = \frac{S_e U_e^3}{S_0 U_0^3} \tag{25}$$

We can make explicit the dependence of the efficiency on geometry by using equations (20a) and (23):

$$C_p(\varphi) = (1 - |\sin\varphi|)\,\cos^2\varphi \tag{26}$$

A condition for maximum efficiency can be obtained by differentiation of equation (26) with respect to $\varphi$ and setting the derivative equal to zero:

$$\frac{dC_p(\varphi)}{d\varphi} = \cos\varphi\,(1 - |\sin\varphi|)(1 \pm 3\sin\varphi) = 0 \tag{27}$$

Equation (27) has only one nontrivial solution:

$$\sin\varphi = \pm\tfrac{1}{3} \quad \cos\varphi = \frac{2\sqrt{2}}{3} \quad \varphi = \pm 0.339837 \ \ (19°28'16") \tag{28}$$

$$\boldsymbol{a' = \frac{3 - 2\sqrt{2}}{3}} \quad \boldsymbol{U_e = \frac{2\sqrt{2}}{3}U_0} \tag{29a}$$

$$\boldsymbol{b = \frac{2}{3}} \qquad \boldsymbol{U_1 = \frac{1}{3}U_0} \tag{29b}$$

$$\boxed{\boldsymbol{C_{p,max} = \frac{16}{27} = 0.592593}} \tag{29c}$$

Equations (29b) and (29c) are identical to (19b) and (19c) and give the theoretical power fraction that can be extracted from an ideal fluid stream. Equation (29a) differs from (19a) and we shall come back on their different meaning later.

Using the value of the *Betz's angle $\varphi$* which was obtained, equation (23) gives:

$$S_e = \frac{\sqrt{2}}{2}S_0 \tag{30}$$

Some misunderstandings in the physical meaning of this result could inadvertently occur and are worth pointing out. The most important implication from the Betz's theory is that in order to extract energy from the stream no turbine can bring the fluid to a total rest. Physically, this necessarily translates in the fact that the effective surface of the turbine which interacts with the fluid must be smaller than the cross-sectional area of the turbine. From a different perspective, the surface element "seen" by the dragged portion of flow corresponding to the absorbed kinetic energy can be expressed as (Fig. 1):

$$dS_e(\varphi) = 2H\tan\varphi\,dx \tag{31}$$

that is

$$S_e(\varphi) = \int_0^L 2H\tan\varphi\,dx = 2LH\tan\varphi = 2S_0\tan\varphi \tag{32}$$

Equating the two expressions (23) and (32) one gets:

$$\frac{1-|\sin\varphi|}{|\cos\varphi|} = 2\tan\varphi$$

which is only satisfied for $\sin\varphi = \pm\frac{1}{3}$ (eq. 28): this represents the **geometric equivalent of Betz condition** on speeds ratio.

It should be noted that during the time the fluid takes to cross the energy extractor thickness with speed $U_1$, $t = (2dx\tan\varphi)/U_1 = 2dx/U_0\cos\varphi$, the latter moves laterally by $t\cdot U_e = 2dx$, that is, exactly the distance which separates two adjacent stripes: the fluid moving axially with speed $U_1$ does not interact with the energy extractor. This is also valid in the finite case, that is, if we replace $dx$ with $\Delta x = L/n$, and represents the geometric reason why there is no induced vorticity in the downstream flow.

Let's now consider a rotating disk. The actuator disk can initially be figured out as a full disk, with its axis parallel to the streamlines, the whole surface of which is perpendicular to the motion of the fluid. The surface of the actuator disk, being orthogonal to the flow, would absorb all the available power ($C_p$=100%), but to rotate and convert the kinetic energy lost by the fluid it must be divided into several circular sectors and each one must be rotated about its radial axis by an angle $\varphi$, with $0 < \varphi < \frac{\pi}{2}$.

In order to recover the perpendicularity between the flow lines and the surface of the actuator disk, we imagine increasing endlessly the number of sectors, by reducing their surfaces that become infinitesimal, and which, therefore, can be considered again orthogonal to the flow, though allowing the fluid to pass through the disk with a residual velocity equal to a fraction of that of the incident flow.

The surface $A$ of the actuator disk perpendicular to the fluid velocity $U_0$ is $A = \pi R^2$, where $R$ indicates the radius of the disk. For a fluid of density $\rho$ the total available power is:

$$P_{tot} = \frac{1}{2}\rho A U_0^3 = \frac{1}{2}\rho\pi R^2 U_0^3 \tag{33}$$

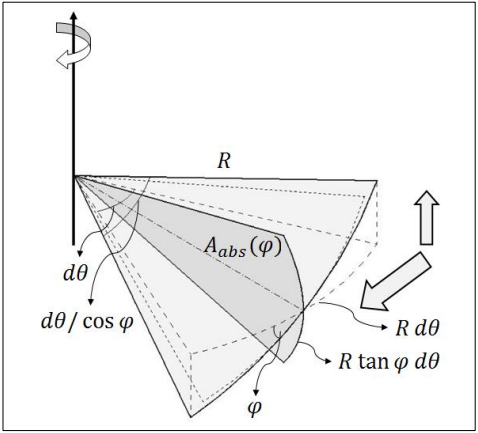

**Figure 2. Geometric proof of Betz's law.**

Let's divide the disk in circular sectors of identical surface $A_1 = \frac{1}{2}R^2\,\theta$, where $\theta$ indicates the arc of the projection of the sectors on the rotor plane and be $\varphi$ the planar inclination about their radial axis (Fig. 2).

The projection of the rotated surface on an orthogonal plane passing through its radial axis is $A_2 = \frac{1}{2}R^2\theta\sin\varphi$. In the limit as $\theta$ approaches zero, the infinitesimal surface of the "stretched" disk blade which sweeps the volume occupied by the rotor can be expressed either by:

$$dA(\varphi) = \frac{1}{2}\left(\frac{dA_1}{\cos\varphi} + \frac{dA_2}{\cos\varphi}\right) = \frac{1}{2}R^2\left(\frac{1+\sin\varphi}{2}\right)\frac{d\theta}{\cos\varphi} \tag{34}$$

or

$$dA(\varphi) = R^2\tan\varphi\cdot d\theta \tag{35}$$

Integrating with respect to $d\theta$ over the whole disk, the actuator disk effective surface "seen" by the absorbable component of the span-wise fluid velocity, $U_{abs}(\varphi) = U_0\cos\varphi$, results:

$$A_{abs}(\varphi) = \int_0^{2\pi}\frac{1}{2}R^2\left(\frac{1+\sin\varphi}{2\cos\varphi}\right)d\theta = \pi R^2\left(\frac{1+\sin\varphi}{2\cos\varphi}\right) \tag{36}$$

For the absorbable power we get:

$$P(\varphi) = \frac{1}{2}\rho A_{abs}(\varphi)[U_{abs}(\varphi)]^3 = \frac{1}{2}\rho\pi R^2 U_0^3\cdot\left(\frac{1+\sin\varphi}{2\cos\varphi}\right)\cos^3\varphi = \frac{1}{4}\rho A U_0^3(1+\sin\varphi)(1-\sin^2\varphi) \tag{37}$$

By differentiating with respect to $\varphi$, one finds that the maximum absorbable power is reached when $\varphi = \pm\arcsin\frac{1}{3}$. Hence:

$$\boldsymbol{U_{abs} = U_0\cos\varphi = \frac{2\sqrt{2}}{3}U_0} \tag{38}$$

$$\boldsymbol{A_{abs} = \frac{\sqrt{2}}{2}A} \tag{39}$$

$$P = \tfrac{1}{2}\rho A_{abs} U_{abs}^3 = \tfrac{1}{2}\rho A U_0^3 \cdot \frac{\sqrt{2}}{2}\left(\frac{2\sqrt{2}}{3}\right)^3 = P_{tot} \cdot \frac{16}{27} \tag{40}$$

that is:

$$\boxed{C_{p,max} \equiv \frac{P}{P_{tot}} = \frac{16}{27}} \tag{41}$$

Note the equivalent meanings of equations (38) and (39) with (29a) and (30), respectively.

This geometric approach to the energy conversion of ideal wind turbines, besides confirming the Lanchester-Betz limit, indicates that the relation (11) between the wake and the rotor induction parameters of the actuator disk model is not a necessary condition for optimal efficiency. Further to this, it allows to understand another important fact. Unlike the linear energy extractor, the thickness in the axial direction of the finite-bladed actuator disk, $r\theta\tan\varphi$, is a function of the radial distance $r$ from the rotation axis; therefore, given the angular separation between two adjacent blades, $\theta$, and their angular velocity, $\Omega$, the fluid with uniform axial speed $U_1 = U_0 \sin\varphi$ will not interact with the energy extractor only if we admit that the angular velocity is inversely proportional to the radial distance, that is, when

$$t \cdot U_1 = \frac{\theta}{\Omega} \cdot U_0 \sin\varphi = r\theta\tan\varphi$$

or

$$\Omega = \frac{U_0 \cos\varphi}{r} \quad \forall\, \theta \neq 0 \tag{42}$$

This demonstrates that the assumption of annular independency, or the assumption of uniformity for a rotating energy extractor, is only valid when $\theta$ tends to zero, which is equivalent to assume the disk has an infinite number of **locally parallel** rotor blades, but it also suggests how this is a purely geometrical aspect that can be eliminated with a topologic transformation.

## 4 The Actuator Cylinder

The geometrical derivation of the Betz law allows us to introduce a conceptual model which will be useful to show the generality of the aerodynamic aspects of energy conversion of an ideal fluid in uniform motion. Consider the following identity:

$$\frac{16}{27} = \left(\frac{\sqrt{2}}{2}\right) \times \left(\frac{2\sqrt{2}}{3}\right)^3 = \sin\frac{\pi}{4} \times (\cos\varphi)^3$$

where $\frac{\pi}{4}$ and $\varphi = \arcsin\frac{1}{3}$ are the *two Betz's angles*. From Betz's law of the maximum absorbable power $P_{abs}$ we have:

$$P_{abs} = P_{tot} \cdot \frac{16}{27} = \tfrac{1}{2}\rho S v^3 \cdot \frac{\sqrt{2}}{2}\left(\frac{2\sqrt{2}}{3}\right)^3 \equiv \tfrac{1}{2}\rho S_{abs} v_{abs}^3$$

where we now assume:

$$S_{abs} = S \cdot \sin\frac{\pi}{4} \qquad v_{abs} = v \cdot \cos\varphi.$$

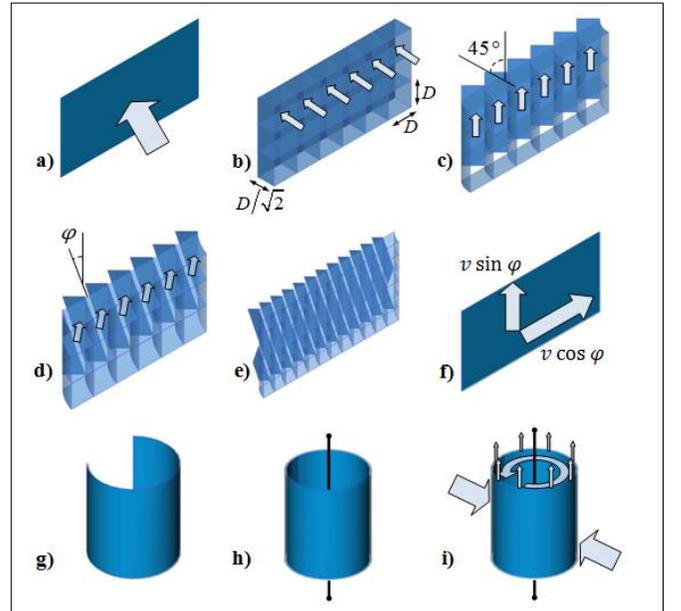

Let's take a rectangular surface perpendicular to the flow (Fig. 3a). Imagine dividing the fluid front ideally into square cells by means of a grid of flow conduits with edge $D$ and depth $D/\sqrt{2}$ (Fig. 3b). Now, bend the end of the flow conduits horizontally so that they direct the streamlines to the right by an angle of 45°. At the same time, we keep the original surface perpendicular to the flow lines by dividing it into vertical stripes and rotating them of 45° accordingly (Fig. 3c). We then bend the conduits vertically in a way to direct the streamlines upward by an angle $\varphi$ of 19°28'16". Once again, we keep the original surface perpendicular to the flow lines by inclining the stripes from vertical position to 90° − $\varphi$ (Fig. 3d).

Like the actuator disk, imagine increasing endlessly the number of cells, by reducing the edge $D$ and grid thickness that become infinitesimal – as well as the surface of the stripes – keeping finite the deflection angles $\varphi$ and 45° (Fig. 3e). We notice that the original surface still absorbs the 100% of the incoming power, but now the flow action can be regarded as divided into two parts: a horizontal component, $v \cdot \cos \varphi$, and a vertical component, $v \cdot \sin \varphi$, both "seeing" a reduced surface: $S \cdot \sin \frac{\pi}{4}$ (Fig. 3f). Finally, consider rolling up the surfaces to form a 2-layer cylinder and to constrain the inner surface in such a way that it can rotate about its axis, while keeping fixed the outer, deflecting, surface (Fig. 3g, h).

**Figure 3. Actuator Cylinder construction.**

By construction, the component of the velocity orthogonal to the axis, $v_{abs} = v \cos \varphi$, is always perpendicular to the reduced surface $S_{abs}$ of the *Actuator Cylinder*, which, in the absence of friction and inertia, will rotate with a tangential velocity equal to it, while the portion of fluid with speed $v_{out} = v \sin \varphi$ never meets the inner surface and will continue undisturbed its motion in axial (vertical) direction (Fig. 3i).

## 5 The complementary Betz's theory

Until now the Betz law has been analytically demonstrated only for HAWTs because it was not immediate how to produce an equivalent theorem for an orthogonal axis. One possibility could be that of splitting the stream tube into two parts: one passing through the front and the other passing through the rear of an actuator cylinder in such a way that the torque exerted at any point is matched by an equal and contrary momentum of the diametrically opposed flow.

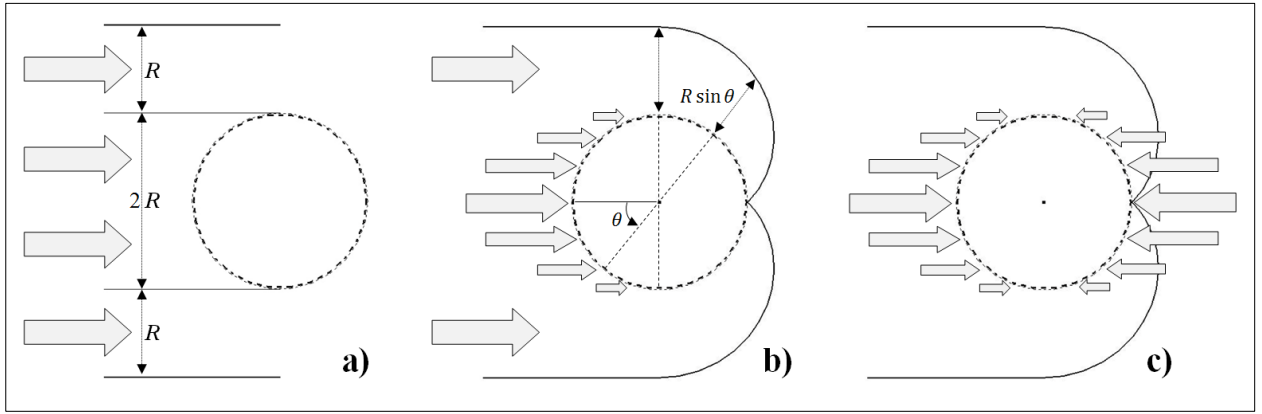

**Figure 4. Leeward fluid collector.**

To achieve this, let's take a stream tube which is large twice the diameter of the actuator cylinder, i.e., let's double the available power (Fig. 4a).

Being the exposed surface of the actuator cylinder not flat, the flow action depends on the angle of incidence $\theta$. Consider collecting the portion of flow at the two flanks of the actuator cylinder and gradually deflecting the streamlines towards the leeward surface, by narrowing the cross-sectional area of a wind guide according to the complementary angle of the diametrically opposite angle of incidence $\theta$ (Fig. 4b). Hint: $\cos \theta = \sin(90° - \theta)$.

The overall effect will be that of exactly replicating on leeward side the wind action on the exposed surface of the actuator cylinder by means of this ideal fluid collector (Fig. 4c). The tunnel effect assumption is justified by the fact that we are considering an uncompressible, irrotational fluid without viscosity or heat exchange with the ideal smooth surfaces.

By indicating with $H$ the height of the actuator cylinder, the total available power $P_{tot}$ is:

$$P_{tot} = \frac{1}{2}\rho(4R \cdot H)v^3 \tag{43}$$

For the absorbable power $P_{abs}$ we have:

$$P_{abs} = \frac{1}{2}\rho\left(2\int_{-\pi/2}^{+\pi/2} R\cos\theta \, d\theta \cdot H \sin\frac{\pi}{4}\right)(v\cos\varphi)^3 = \frac{1}{2}\rho\left(2\sqrt{2}R \cdot H\right)\left(\frac{2\sqrt{2}}{3}\right)^3 v^3 \tag{44}$$

From the ratio of these two expressions, we obtain Betz's result:

$$\frac{P_{abs}}{P_{tot}} = \frac{\left(2\sqrt{2}\right)^4}{4 \cdot 3^3} = \frac{16}{27}$$

where, however, this time we started from a flow no longer parallel but orthogonal to the rotation axis of the ideal energy converter used, the actuator cylinder.

It has been often pointed out that Betz's theory demonstration is formally true only for an actuator disk having an infinite radius. The same can be argued also for the complementary Betz's theory, which holds for an actuator cylinder of infinite radius $R$ – and of infinite height as well, because ideally $H = \pi R/\sqrt{2}$. Anyway, in both cases the numerical result of the maximum allowed efficiency of fluid turbines can be assumed valid by induction.

<p style="text-align:center">***</p>

It is evident that both the Actuator Disk (AD) and the Actuator Cylinder (AC) are ideal concepts useful for calculating the theoretical Betz limit of 59.3%, but, together, they allow us to state which are the inescapable hypotheses to achieve the highest mechanical efficiency of <u>any</u> turbine:

1. *Constancy of the angle of incidence between flow lines and blade surface during rotation*
2. *Perpendicularity of incidence of flow lines on blade surface*
3. *Uniformity of distribution of the flow action over the whole blade surface*
4. *Parallel motion of flow lines with respect to the axis of rotation, i.e.*
5. *No dispersion of flux tube in crossing the rotor (continuity).*

The first assumption alone should not leave any doubt on which is the reason why VAWTs cannot compete in terms of efficiency with HAWTs, even neglecting known considerations about passive phase of upwind travel or absence of lift on the blades which, to variable extent, affect all vertical axis wind turbines.

<p style="text-align:center">*The "perfect turbine" cannot be a VAWT, as it cannot be a HAWT either!*</p>

There have been many attempts to increase the efficiency of real turbines, typically by using air conveyors surrounding the rotor, where the general idea is to create a low-pressure region to drive a greater airflow on the blades. In all cases, it was found that an equivalent turbine having a swept surface equal to the area intercepted by the conveyor had a better performance than the one under examination. The rather obvious conclusion is that the Betz's theory remains valid and the flux tube to be considered has a diameter closer to that defined by the conveyor than to that of the inner rotor[7]: there is no way to bypass the energy balance of the airflow taken at great distance upstream and downstream from the turbine, and all the cited hypotheses confirm themselves as mandatory.

Is it possible in principle to realize real turbines which are able to respect all these hypotheses and, therefore, to approach the efficiency limit? To answer this question, we analyze an ideal turbine in the form of a *finite actuator cylinder*. It may be considered the equivalent of a 3-blade mill as compared to the actuator disk.

## 6 Conceptual design of a 3-D Actuator Cylinder

This section describes the conceptual design used for the proposed case study and the geometrical construction of its constituent elements, with the aim to ease the subsequent analysis of its aerodynamic properties.

The simplest finite actuator cylinder may be composed of two identical truncated-cone surfaces, which represent the lower and upper base of the structure delimiting the height of the stream tube, plus a series of intermediate sections with the same inclination of the two bases, same inner radius $r$, and external radius $R$ equal to half of that of the bases, $2R$, arranged coaxially and equidistant from each other (Fig. 5a).

Bases and sections have an inclination with respect to the horizontal plane of an angle $\beta$ such that the gradient at 45° with respect to the radial direction equals the tangent of the angle $\varphi$ of 19°28'16", or:

$$\tan\beta = \tan\varphi \cos^{-1}\frac{\pi}{4} = \frac{1}{2} \qquad \beta = 26°33'54".$$

These surfaces are rigidly fixed to each other by a set of $N$ vertical walls, whose curved section represents an arc of cycloid of parametric equation:

$$x = \frac{D}{4}(\theta + \sin\theta), \ y = \frac{D}{4}\cos\theta \ \text{ with } 0 \le \theta \le \frac{\pi}{2}, \tag{45}$$

of length $D/\sqrt{2}$, where $D = 2R\sin(\pi/N)$, radially arranged at regular distances around the inner edge of radius $r$ and extending up to a distance $R$ from the axis of symmetry of the AC (Fig. 5b).

The set of conical surfaces and vertical walls defines the plurality of flow conduits, of outer section $D^2$, which delimit the cylindrical rotor space at the center of the AC.

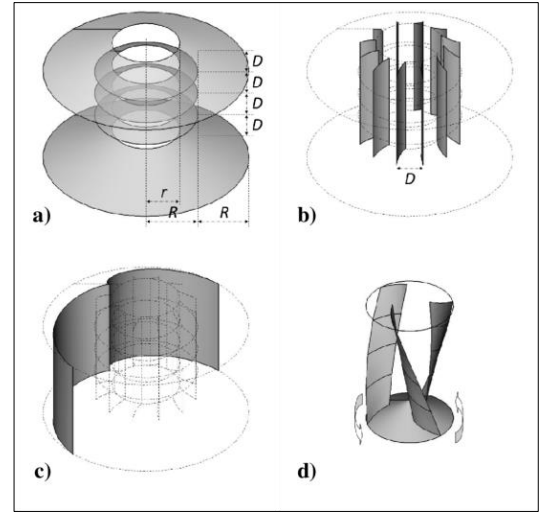

**Figure 5. Constituent elements of the AC concept: a) bases and sections; b) vertical walls; c) collector; d) rotor.**

Differentiating the equation (45), we find that the velocity of the flow along the trajectory through each conduit forms in the horizontal plane an angle $\alpha(\theta) = \arctan\left(\frac{dy}{dx}\right) = \frac{\theta}{2}$ with respect to the initial direction $\alpha(0) = 0$. Therefore: $\alpha\left(\frac{\pi}{2}\right) = \frac{\pi}{4}$. On a horizontal plane, the flow lines inside each conduit enter the rotor space with a constant inclination of 45°. On a vertical plane, the flow lines enter the rotor space with a constant inclination $\varphi =$19°28'16" in the axial direction.

The blades will have an axial torsion equal to the complementary of the angle $\varphi$ and the surface is oriented at 45° with respect to the radial direction. In fact, the surface perpendicular to the flow lines that best describes the ideal shape of the blade in the horizontal plane is expressed by the parametric equation:

$$x = r\theta, \ y = -r\theta \tan(1-\theta) \tag{46}$$

where $r$ is the radius of the cylindrical inner rotor space, of length:

$$l = R \cdot \cos\frac{\pi}{N}\sin\varphi \tag{47}$$

*Demonstration.* Consider decomposing the planar (horizontal) component of the injected air velocity, $v_i^{(h)}$, into a tangential subcomponent, $v_\tau$, and a radial one, $v_\parallel$. Because of the geometric constraint, the two sub-components have the same modulus: $v_\tau = v_\parallel = v_i^{(h)} \cos\frac{\pi}{4}$. Neglecting frictional forces, the angular velocity at which the zero-inertia rotor will be forced to rotate is $\omega = v_\tau/r$ . In a time $t$ the blade rotates by an angle $\theta = \omega t = (v_\tau/r)t$. Along the radial axis, at the same time $t$, air travels a distance $y = v_\parallel t = v_\tau t = r\theta$. Starting from the initial incidence point O(0, 0), the air that moves in the radial direction will hit the blade if, after rotating by an angle $\theta$, the blade intersects the radius at point P(0, $r\theta$). This means that when the edge of the blade is at the origin O, its generic point of ordinate $y$ must have an abscissa given by $x = -(r - y)\tan\theta$. Thus, the equation that describes the ideal blade profile is $x = -(r - y)\tan\frac{y}{r}$, corresponding to the parametric form (46).

The two bases support at the bottom and at the top, a mobile element called collector, free to swing around the AC to set itself downwind, ideally delimited by an external vertical surface of semicircular section and an internal vertical surface, facing the AC, whose section is composed of two specular arcs of a cardioid, which starting from the two ends of the semicircle of radius $2R$ meet in a central cusp placed at a distance $R$ from the axis of the structure (Fig. 5c).

The shape of the leeward collector originates by imposing that the distance, $R \sin\theta$, between its surface and the cylinder of radius $R$, varies from a minimum for $\theta = 0°$ to a maximum for $\theta = 90°$, by reason of an angle of incidence $\theta$ between the direction of the outside air and the windward surface of the AC.

Finally, the hollow rotor is devoid of a central shaft. The inner region of the blades is occluded at the bottom by a right cone basement, integral with the blades, which completes the truncated cone lower base, while its upper represents the air discharge opening (Fig. 5d).

# 7 Simplified aerodynamic analysis

We now verify with a direct calculation that, for its particular geometry, the efficiency of this interesting crossflow design turbines approximates the limit value, assuming ideal surfaces and neglecting turbulence in the air flow.

Let be $D = 2R \sin(\pi/12)$ the height of each flow conduit or sector, having chosen for this exemplary embodiment twelve vertical walls to delimit the rotor space ($N = 12$). As a function of this scale factor $D$ the following proportions apply:

$$R = \frac{D}{2}\left(\sin\frac{\pi}{12}\right)^{-1} = \frac{\sqrt{6}+\sqrt{2}}{2}D \approx 1.932 \cdot D \tag{48}$$

$$r = \sqrt{R^2 - \left(\frac{D}{4}\right)^2} - \frac{D}{4}\left(1 + \frac{\pi}{2}\right) = \left[\sqrt{2 + \sqrt{3} - \frac{1}{16}} - \frac{1}{4}\left(1 + \frac{\pi}{2}\right)\right]D \approx 1.273 \cdot D \tag{49}$$

Considering $s$ superimposed sector orders, the overall effective area of air interception from the AC is:

$$A_e = 4R \cdot sD = 2s\left(\sqrt{6} + \sqrt{2}\right)D^2 \tag{50}$$

Indicating with $v_e$ the external, or entering, wind speed, and with $\rho$ the air density, the total available power $P_e^{max}$ is expressed by:

$$P_e^{max} = \frac{1}{2}\rho A_e v_e^3 = s\left(\sqrt{6} + \sqrt{2}\right)\rho D^2 v_e^3 \tag{51}$$

The internal area of incidence $A_i$ for each flow conduit is:

$$A_i = D^2 \cos\frac{\pi}{4}/\cos\varphi = \frac{3}{4}D^2 \tag{52}$$

In a conduit whose vertical side walls are two identical cycloid arcs, converging according to an angle of 30° (360°/12), the incoming air is subjected to an acceleration by tunnel effect inversely proportional to the narrowing of the cross-section area. In-dicating with $v_i$ the internal, or injection, velocity, assuming a laminar flow and being $D^2$ the intake area of each flow conduit, we obtain that the injection air velocity is greater than the external wind speed by a factor:

$$v_i = \frac{D^2}{A_i} v_e = \frac{4}{3} v_e \qquad (53)$$

The sectors contribute to the overall power in a different extent according to the orientation of their inlet sections relative to the incident wind direction; without loss of generality, for all the conduits that surround the rotor, we can consider the ensuing three possible angulations: 75°, 45° and 15° (or 90°, 60°, 30° and 0° for the other limit position). Since the leeward sectors receive an amount of air equal to the windward sectors by virtue of the collector geometry, the flow intercepted by the two frontal sectors, placed at 75° with respect to the wind direction, proportional to sin 75°, is equal to that of the two diametrically opposite sectors, in correspondence of the cusp of the collector; four sectors intercept a flow proportional to sin 45°, the remaining four sectors proportional to sin 15°. By summing all the contributions of the conduits of the AC, the total power within the rotor space is expressed by:

$$P_i^{max} = s \cdot \frac{1}{2} \rho A_i \cos\frac{\pi}{12} \cdot 4 \left[ \sin\frac{5\pi}{12} \left( v_i \sin\frac{5\pi}{12} \right)^3 + \sin\frac{\pi}{4} \left( v_i \sin\frac{\pi}{4} \right)^3 + \sin\frac{\pi}{12} \left( v_i \sin\frac{\pi}{12} \right)^3 \right] = s \frac{9}{4} \rho A_i v_i^3 \cos\frac{\pi}{12} \qquad (54)$$

By replacing the values of $A_i$ and $v_i$ given by (52) and (53), and taking into account equation (51), we get:

$$P_i^{max} = s \frac{9}{4} \rho A_i v_i^3 \cos\frac{\pi}{12} = s \frac{9}{4} \rho \cdot \frac{3}{4} D^2 \cdot \left( \frac{4}{3} v_e \right)^3 \frac{\sqrt{6}+\sqrt{2}}{4} = s(\sqrt{6}+\sqrt{2})\rho D^2 v_e^3 = P_e^{max} \qquad (55)$$

As expected, the available power within the rotor space is equal to that intercepted by the collector, in the assumption of laminar flow and smooth surfaces.

To calculate the power absorbable by the rotor, we decompose the injection velocity $v_i$ in its horizontal component, $v_i^{(h)}$, and its vertical component, $v_i^{(v)}$, which are, respectively, $v_i^{(h)} = \cos\varphi \cdot v_i = \frac{2\sqrt{2}}{3} v_i$ and $v_i^{(v)} = \sin\varphi \cdot v_i = \frac{1}{3} v_i$.

The work orthogonal to the rotation axis done on the blades is only due to the horizontal component $v_i^{(h)}$. By replacing the latter into equation (54) and considering the perpendicular projection of the inner surface with respect to the axis, $A_i \cos\frac{\pi}{4}$, the absorbable power results:

$$P_i = s \frac{9}{4} \rho \left( A_i \cos\frac{\pi}{4} \right) (\cos\varphi \cdot v_i)^3 \cos\frac{\pi}{12} = \cos\frac{\pi}{4} (\cos\varphi)^3 \cdot P_i^{max} = \frac{\sqrt{2}}{2} \left( \frac{2\sqrt{2}}{3} \right)^3 \cdot P_i^{max} = \frac{16}{27} \cdot P_i^{max} \qquad (56)$$

According to the assumption that the power corresponding to the horizontal component of the injected air be entirely absorbed, the theoretical efficiency of the turbine coincides with the Betz's limit.

It should be emphasized that this remarkable result is only possible with the particular choice of the Betz angle $\varphi$ of 19°28'16" for the inclination of the conical surfaces. Any other angle $\varphi'$ such that $\sin\varphi' \neq \frac{1}{3}$ would lead to a theoretical value of $C_P$ less than 59.3%, just as in the case of the actuator disk when the Betz condition is not respected, i.e., when $\frac{v_{out}}{v_{in}} \neq \frac{1}{3}$.

The results (55) and (56) relating to the available and absorbable power assume that the rotor be composed of a number of blades equal to the number of deflecting walls that delimit the rotor space. We shall now verify how the lift force acting on the blades along their entire path, or the absence of a resistant phase, allows the use of an impeller with a lower number of blades.

In absence of the rotor, the air injected by each sector, not impinging on any blade, would continue until it intersects the flow exiting the next conduit according to a constant angle and the respective velocities would sum vectorially.

Each sector contributes to increase the tangential velocity of the vortex proportionally to the sine of the angle formed by its own inlet section with respect to the wind direction. Since the ratio of the tangential velocity and the axial velocity of the vortex is constant at every point, it is always possible to choose the height of the AC such that the air entering at the base exits from the top of the rotor space after completing one full revolution.

With this assumption, considering the possible angulations 15°, 45° and 75°, the average injection velocity is expressed by:

$$\bar{v} = \frac{\sin\frac{\pi}{12}\sin\frac{\pi}{12} + \sin\frac{\pi}{4}\sin\frac{\pi}{4} + \sin\frac{5\pi}{12}\sin\frac{5\pi}{12}}{\sin\frac{\pi}{12} + \sin\frac{\pi}{4} + \sin\frac{5\pi}{12}} v_i = (\sqrt{6}-\sqrt{2}) v_e \qquad (57)$$

Starting at the level of the lower base and proceeding in the direction of counterclockwise rotation, from simple geometric considerations we have for the velocities of the vortex in correspondence of the different sectors:

| | | | |
|---|---|---|---|
| $v_1^2 = \bar{v}^2$ | $v_1 = 0.78 \cdot v_i$ | $v_7^2 = \frac{9}{4}\left(\frac{7}{2} - \sqrt{3}\right)v_i^2$ | $v_7 = 1.99 \cdot v_i$ |
| $v_2^2 = \frac{9}{8}(3-\sqrt{3})v_i^2$ | $v_2 = 1.19 \cdot v_i$ | $v_8^2 = \frac{9}{16}(10-\sqrt{3})v_i^2$ | $v_8 = 2.16 \cdot v_i$ |
| $v_3^2 = \frac{9}{4}v_i^2$ | $v_3 = 1.50 \cdot v_i$ | $v_9^2 = \frac{45}{8}v_i^2$ | $v_9 = 2.37 \cdot v_i$ |

$$v_4^2 = \frac{45}{16}v_i^2 \qquad v_4 = 1.68 \cdot v_i \qquad\qquad v_{10}^2 = \frac{99}{16}v_i^2 \qquad v_{10} = 2.49 \cdot v_i$$

$$v_5^2 = \frac{9}{8}\left(\frac{9}{2} - \sqrt{3}\right)v_i^2 \quad v_5 = 1.76 \cdot v_i \qquad\quad v_{11}^2 = \frac{9}{16}\left(13 - \sqrt{3}\right)v_i^2 \quad v_{11} = 2.51 \cdot v_i$$

$$v_6^2 = \frac{9}{4}\left(\frac{13}{4} - \sqrt{3}\right)v_i^2 \quad v_6 = 1.85 \cdot v_i \qquad\quad v_{12}^2 = \frac{9}{4}\left(\frac{19}{4} - \sqrt{3}\right)v_i^2 \quad v_{12} = 2.61 \cdot v_i$$

In absence of blades, therefore, the velocity increases gradually, passing from one sector to the next.

Said $v_b$ the air velocity before the blade and $v_a$ the velocity after the blade, with $v_a > v_b$, indicating with $A_l$ the area subjected to the pressure gradient, the power due to this lift effect is expressed by the formula:

$$P_l = \frac{1}{2}\rho A_l \bar{v}\left(v_a^2 - v_b^2\right) \tag{58}$$

The section $A_l$ orthogonal to the flow, taking into account expression (47), will be:

$$A_l = l \cdot D = R \cdot \cos\frac{\pi}{12}\sin\varphi \cdot D = \frac{(2+\sqrt{3})}{6} \cdot D^2 \approx 0.622 \cdot D^2 \tag{59}$$

The surface $A_l$ on which the pressure gradient acts is smaller than the area of incidence $A_i \approx 0.75 \cdot D^2$ previously considered because only a fraction of the air coming out from each sector spreads out in the area of influence of the adjacent sector producing an accelerated airflow. In particular, the tip of the blades is not affected by any pressure gradient (same tangential speed on leading and trailing surfaces), which qualitatively explains why $A_l < A_i$.

We can use the expression (58) to calculate the total power due to the lift effect alone, by setting $v_a = v_{n+1}$ and $v_b = v_n$ ($v_0^2 \equiv v_1^2 - v_e^2$):

$$P_l^{max} = \frac{1}{2}\rho A_l \bar{v}[s\sum_{n=0}^{11}(v_{n+1}^2 - v_n^2)] \tag{60}$$

Replacing the value of $A_l$ and the average velocity $\bar{v}$, and developing the summation

$$\sum_{n=0}^{11}(v_{n+1}^2 - v_n^2) = \frac{27}{4}v_i^2 = 3\left(2 + \sqrt{3}\right)\bar{v}^2 = 12v_e^2$$

we get:

$$P_l^{max} = \frac{1}{2}\rho D^2 v_e^3 \frac{2+\sqrt{3}}{6}\left(\sqrt{6} - \sqrt{2}\right)12s = s\left(\sqrt{6} + \sqrt{2}\right)\rho D^2 v_e^3 = P_i^{max} = P_e^{max} \tag{61}$$

The result obtained according to aerodynamic lift considerations perfectly matches the value of the total available power previously calculated and rests on the assumption of the absence of the rotor. In presence of the rotor, air generally impacts a blade before reaching the next sector; the diffusion and the consequent merging of the flow lines involves a number of adjacent sectors which gradually decreases as the number of blades composing the rotor increases, up to vanish completely when the number of blades is equal to the number of sectors, $N$. On the other hand, the available internal power $P_i^{max}$ calculated according to standard considerations assumes, instead, that the rotor is composed of $N$ blades equal to the number of radial sectors; a lower number of blades means that one or more of the terms of equation (54) is zero and, consequently, that the total available power progressively reduces until vanishing in absence of blades. A direct calculation shows that for an intermediate number of blades between 0 and $N$ – provided they are arranged at regular angular distances – the "thrust" contribution to the total power that is missing is compensated in equal measure by the augmented aerodynamic "lift" on the first available blade preceding the absent blade.

## 8 Discussion and conclusions

Starting from an interesting approach to the classical theory and a geometric explanation of Betz law, the complementary Betz's theory has been analytically deduced from a rotor with axis orthogonal to the direction of an ideal fluid in uniform motion. The conceptual design used to demonstrate the generality of the aerodynamic aspects of energy conversion shows the important role played by the Betz's angles in extending the theory of flow turbines from a flat, two-dimensional, vision to a three-dimensional one and provides a more general understanding of the problem.

The achieved result, in particular the possibility of obtaining constant yield and maximum efficiency with a finite topology such as that of a 3-D actuator cylinder, suggests how this is in principle feasible with a real turbine with axis perpendicular to the wind direction.

As shown in Fig. 6, the freely swinging collector PCQ, by virtue of its shape and constraints, produces air streams in the leeward flow conduits similar to those of the flow conduits exposed to the wind (white arrows). This contributes to sustain a cyclonic circulation of air in the rotor space, enhancing its cylindrical symmetry and concentricity with the rotor axis (inner dotted arrows) and eliminating any braking action on the blades.

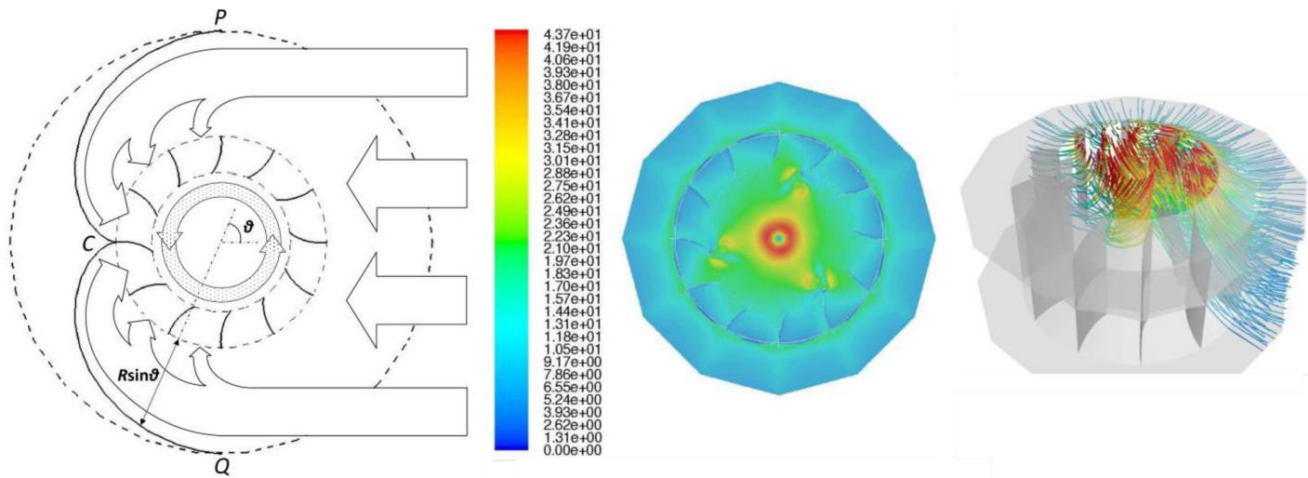

**Figure 6. Collector geometry and analysis of the induced vortex.**

Wind air intercepted by the two wings of the mobile collector is gradually deflected and directed toward the leeward surface of the cylindrical central part of the structure. A stream of identical flow rate is directly intercepted by the windward surface. The short flow conduits guide these two airstreams into the inner rotor space, where they assume an ascending swirling motion of cylindrical symmetry. A rotor of any shape and size sweeps a full cylindrical volume (*Savonius*, HAWT) or a tubular volume (*Darrieus*, *Gorlov*). Therefore, a steady cyclonic flow, confined in the inner rotor space, regular and concentric with the axis of rotation of the turbine, allows maintaining a constant angle of incidence between the flow lines and the surface of the blades, as imposed by the **first hypothesis** introduced in §5.

The blades have an axial torsion equal to the complementary of the angle $\varphi = 19°28'16''$ and the surface is oriented at minus 45° with respect to the radial direction. The jets exiting the guiding conduits are constantly slanted upward by an angle $\varphi$ and deflected by 45° to the right (tribute to *Coriolis*). Therefore, the flow lines hitting the blades impinge perpendicularly onto their surfaces during the complete revolution as required by the **second hypothesis**. Moreover, being the streamlines exiting the flow conduits slightly convergent, the peripheral surface of the blades near the jets is convex to be ideally perpendicular to this *primary* injection flow.

By opting for a number of blades composing the impeller lower than the number of flow conduits, the air injected therein that does not meet a blade constitutes the *secondary* injection flow, the energy of which boosts the vortex, ensuring the whole surface to be subject to a constant lift effect, generated by the concordant movement of the air that precedes the blades, in compliance with the **third hypothesis**, and that the impeller does not encounter resistance during its full rotation.

Contrarily to traditional VAWTs, the air thrust, corresponding to the horizontal component of the injection speed, acts on the *convex* side rather than on the *concave* side of the blades. The reason for this lies on the fact that, while in common generators the intent is to reduce the resistance that the blades meet in the upwind travel, in the turbine under examination the passive phase is absent, and attention may focus on optimizing the energy transfer from the flow to the rotor.

In the assumption of a zero-inertia impeller with no friction or exchange of heat, the whole energy associated to the vertical component of the injection velocity is not absorbed by the rotor and guides the axial exhaust flow (*Coandă effect*), as contemplated by the **fourth hypothesis**.

Notice that, despite the apparent paradox, the case study turbine is not a VAWT. The distinction between horizontal and vertical axis turbines is commonly referred in an "anthropic" way to the ground because the wind blows parallel to the earth surface, but more correctly it should be referred to the main direction of the streamlines. In this model the flow that moves the rotor crosses it in axial direction, as in all HAWTs

Of course, in presence of a real fluid and rotor, not all of the primary and secondary flows energy is transferred to the blades. The unavoidable turbulence feeds an innermost vortex, in proximity of the rotation axis, in a void space not swept by the blades, wherein the fluid reaches very high speeds. This low-pressure inner region ensures the confinement of the flux tube during the crossing of the rotor required by the **fifth hypothesis**. Moreover, as in a whirlwind, the low-pressure core would help reaching a steady state whereat incoming wind air is attracted by the turbine and slightly compressed before hitting the blades, positively affecting the intrinsic efficiency. In other words, the slowdown of the stream caused by the presence of the obstacle represented by the turbine structure is expected to be reduced with respect to any other type of wind generators of identical interception area.

The analyzed model may be qualified as a "cyclonic-flow turbine" because the rotor is driven by a self-sustaining rotational movement of the fluid induced by a *stator* – that is by what we have called the finite actuator cylinder – and which proceeds autonomously towards one of the ends of the inner rotor space of the cylinder.

In the cyclonic flow turbine, the solution adopted is to exploit the natural predisposition of a fluid to rotate in presence of a central outflow and pressure drop. It behaves like an axial flow wind turbine, such as windmills, though offering the practical characteristics of vertical axis turbines.

The expected yield is close to the efficiency limit due to the absence of passive phase. Air acts uniformly and always orthogonally to the blades, without flow dispersion. The pressure gradient between the inner void space and the outside of the AC structure and a continuous lift effect on blades contribute to the power output.

A constant yield on a larger range of wind speeds and low startup velocity would appear to be other significant characteristics of this model turbine.

We count on the present theoretical work to stimulate the discussion within the community and address further research and proof of concept towards the improvement of numerical design tools and the development of future generations of wind turbines.

## Acknowledgements

I am grateful to the D.I.M.E., the Department of Mechanical and Energy Engineering of the University of Genoa, Italy, for the stimulating discussions and the preliminary CFD simulation provided by Prof. Pietro Zunino and his collaborators. A special thanks go to Prof. Ernesto Benini of the Department of Industrial Engineering of the University of Padua, Italy, for his early interest in my work and support for a deeper numerical analysis.

## Declaration of Conflicting Interest

The author declares that there is no conflict of interest in publishing this article. An international patent has been filed concerning possible device design and industrial application suggested by present conclusions with the aim to avoid private speculation and access limitation to the possible positive implications of this theoretical work.